\RequirePackage{fix-cm}
\documentclass{svjour3}                     
\smartqed  
\usepackage{graphicx}
\usepackage{CJKutf8}
\usepackage{multirow}
\usepackage{booktabs}
\usepackage{rotating}
\usepackage{geometry}
\usepackage[colorlinks,linkcolor=black]{hyperref}

\begin{document}

\title{Evaluating Online Public Sentiments towards China
}
\subtitle{A Case Study of English and Chinese Twitter Discourse during the 2019 Chinese National Day}


\author{Xu Yekai         \and
        He Qingqian \and
        Ni Shiguang*  
}


\institute{Xu Yekai \at
              Big Data Lab, Tsinghua-Berkeley Shenzhen Institute, Tsinghua University, Shenzhen, China \\
			  Tsinghua International Relations Data \& Computing Lab, School of Social Sciences, Tsinghua University, Beijing, China
           \and
           He Qingqian \at
              Tsinghua International Relations Data \& Computing Lab, School of Social Sciences, Tsinghua University, Beijing, China
		   \and
		   Ni Shiguang  \at
		   Graduate School at Shenzhen, Tsinghua University, Shenzhen, China  \\
              \email{ni.shiguang@sz.tsinghua.edu.cn}           
		   \and
		   Associated data repository can be found here: \url{https://github.com/JaimeLaVie/National_Day_Public_Opinion_Mining/tree/master/data_0929_1003}. The entire dataset is too large and cannot be uploaded, so a sample dataset of a critical period is provided.
}

\date{Received: date / Accepted: date}

\maketitle

\begin{abstract}
As the Internet gradually penetrates into people’s daily lives and empowers everyone to demonstrate and exchange opinions and sentiments online, individual citizens are increasingly participating in the agenda-setting of public affairs and the design and implementation of official policies. The current study describes an approach to analyze online public sentiments using social media data and provides an example of Twitter discourse during the 2019 Chinese National Day. Over 300,000 tweets were collected between Sept 30 and Oct 3, and a hybrid method of SVM and dictionary was applied to evaluate the sentiments of the collected tweets. This method avoids complex structures while yielding an average accuracy of over 96\% in most classifiers used in the study. The results indicate alignment between the time of National Day celebration activities and the expressed sentiments revealed in both English and Chinese tweets, although the sentiments of the two languages tend to be in different directions. The sentiment of tweets also diverges from nation to nation, but is generally consistent with the country’s official relations with China to varying degrees. The linguistic features of the tweets suggest different concerns for Twitter users who have different sentiments towards China. At last, possible directions for further studies are indicated.
\keywords{Attitudes towards China \and Chinese National Day \and Twitter data \and Sentiment analysis \and Hybrid method \and Temporal, spatial and lingual characteristics}
\end{abstract}

\section{Introduction}
\label{sec:1}
\label{intro}
In modern age, public opinion has long been playing an important role in domestic politics, international relations and the associated policy making. An illustrative case is the U.S. withdrawal from Vietnam in 1970s, which could be partially attributed to the massive demonstrations. Existing research also proves an alignment between news’ opinions and international events and relations. [1] In the last two decades, online public opinion has proved itself to be as influential as, if not even more than, the opinion delivered by traditional media. A recent example is the controversial tweet by Daryl Morey, which later led to online conflicts and affected the China-US relations. The Ministry of Foreign Affairs of China gave an official response to the problem, and politicians including Hillary Clinton and Mitchell McConnell also took advantage of the incident to promote their political claims. On the other hand, cyberspace empowers non-state and even traditionally non-political actors to join the discussions of political issues, helping constructing social reality.[2] Considering the highly open and interconnected nature of the Internet, millions of data with all kinds of attitudes towards different political entities are being generated every minute, making online public opinion an indispensable factor in analyzing politics and policies. 

The 70th anniversary of the foundation of the People’s Republic of China, which has attracted worldwide attention, provided a rare opportunity for analyzing the online sentiments towards China. Large amount of messages related to Chinese National Day were published on Twitter, one of the leading social media globally. The volume of such data, for the first time in Chinese history, enables the tracking and observation of the expressed sentiments of all kinds of people around the world during this kind of significant occasions. Thus, this research attempts to gain a comprehensive view of China on the Internet through Twitter data and natural language processing algorithms. We collected Chinese National Day-related tweets around October 1st and conduct sentiment analysis to each and every tweet within the dataset to obtain temporal, spatial and linguistic insights into the overall sentiment towards China on one of the mainstream social platform. We first introduce the collecting and labeling process of the tweets, and examine the validity of our dataset and labels. Then, we present methods used in this study. After that, through experiments, we show the competency of the methods, as well as the analysis with regards to time, countries and language use. In the last, we conclude our research and highlights the significance of the experimental results.

\section{Previous Works}
\label{sec:2}
It has become an emerging field in recent years to apply social media data for analyzing political and international relations problems. The key technique used text-based is opinion mining.[3] This section reviews some of the important works.

Whether social media data can be applied to predict the result of political elections and referendums is one of the hot topics. Tsirakis et al. did a primitive attempt by analyzing the public opinion expressed through Twitter during a debate before the 2015 Greek election.[4] Tsakalidis et al. applied a semi-supervised approach to conducting a daily analysis of the voting intentions of over 2,000 Twitter users during the 2015 Greek election.[5] Jungherr et al. collected 3 months of Twitter data before the 2013 German federal election to predict the public opinion and compared the results with human polls and the final election outcomes. Both SVM and simple mentioning methods were used in this paper. The authors concluded that social media is more an indicator of attention rather than political support.[6] Bovet et al. used statistical physics combined with machine learning methods to investigate public opinion during the 2016 US presidential election. The results were found to be remarkably accurate compared with the New York Times National Polling Average.[7] Unlike forecasting the result of political elections, Lopez et al. utilized over 23 million Tweets about Brexit to “nowcast” (real-time monitor) the public opinion about Brexit referendum. And this paper differs from the previous one in that it regards social media data as informative, or at least strongly supplementary to polls, for forecasting and nowcasting political support in elections. [8]

Despite elections, social media data was also used to analyze sentiments towards certain political entities. Gull et al. used support vector machine (SVM) to study opinionated English tweets towards Pakistani political parties. [9] Gong et al. applied convolutional neural networks (CNN) to study the sentiments of China-related posts on Reddit from 2007 to 2015 and discovered that these posts showed a “skewed spread” characteristic, with negative information occupying the dominant position and significantly overweighting positive opinions. [10] However, both researches neglects the non-English texts, which makes results somehow unrepresentative. This is particularly noteworthy for [9] due to the fact that the majority of the Pakistani population mainly speak local languages rather than English.

There are several studies utilizing non-English social media data for analyzing sentiments. Jamal et al. analyzed the anti-Americanism and anti-interventionism in the Middle East using Twitter texts written in Arabic. The main point of the article was to answer the question of whether the prevalent anti-Americanism and animus in the Middle East were targeted at the U.S. itself or towards the impingement of the United States on other countries.[11] A similar study is [12], which focused on the views of Chinese Weibo users towards the United States and several other neighboring nations. Weibo discourses were collected as datasets. This paper found that Chinese views of the United States were deeply ambivalent: positive to American technology advances while negative to American foreign policy. In addition, Di Giovanni et al. used Italian tweets to predict users’ political inclinations.[13] However, albeit Twitter data were carefully observed in these studies for political analysis, the utilization of social media data was rather primitive. They lack in-depth study of the exact meaning and sentiment tendency of the tweets. Therefore, the coarsely processed twitter data were not able to fulfill their potential. Meanwhile, although these researches took advantage of non-English data, they are still limited to a single language, and other languages used in the studied areas were overlooked.

Social media data can also provide insights into political relations. For example, Barnett et al. developed an approach to analyze characteristics of international relations based on the co-occurrence of nations' names in short texts extracted from Facebook and Weibo. [14] Kumar et al. used Twitter data (including hashtags, followers, retweet, mention, etc.) and network analysis to investigate relations between Latin American political organizations. [15]

A similar research to this work were conducted by Chambers et al. They applied Twitter data to model relations between nation-states using sentiment analysis, which is a synthesis of the above mentioned three categories. Compared with traditional polling data, data acquisition from social media can be time-saving and low-cost. They picked out Tweets containing country names, and then filtered the data to remove irrelevant tweets like restaurants, sports meetings, etc. The remaining tweets were used to calculate the aggregated nation-nation sentiments. This study provided a temporal result for international relations and verified it with human polls. Therefore, it proved the applicability of utilizing media text data for international relations studies. [16] However, all the non-English tweets were still filtered out in this work. Thus, its results could only represent the English-speaking population. On the other hand, this work merely cared about the overall general relations between different nations. The influences of certain events to international relations and the relationship of specific nation pairs are not fully studied. Another similar study investigated user demographics and topic diversity of Twitter discourse during the Global Landscapes Forum (GLF) 2018 in Nairobi. The study found that convening the GLF in Africa enhanced the voice of Africans compared to the previous GLF in Germany.[17] Whereas, albeit it focused on a specific event, the core of the paper is demographic differences in topic-level engagement between different races, genders and ages. The study provides little implication for the international politics aspect and the relationship between relevant nations.

There exist researches that intend to introduce multilingual or multimodal analysis. Rashkin et al. established multilingual connotation frames using existing parallel corpus with automatic word-alignment for English and 10 other European languages to analyze the targeted sentiments of users with diverse language backgrounds. [18] Over 1.2 million tweets were collected from Twitter for the study and LSTM was utilized to predict future sentiments with previous sentiments data. This paper verified the applicability of exploiting cross-language corpus to retrieve and predict targeted sentiments and public opinions of different nations, but more fine-grained and specific studies are still needed. It is also possible to conduct opinion mining from multimodal social media data. Fang et al. analyzed texts and photos simultaneously to realize aspect-opinion retrieval. [19] In addition, Peña-Araya et al. and Fernando et al. independently developed visualization tools to study the geo-temporal contexts in Twitter related to history, politics, and international relations. [20, 21]

This paper differs from the above mentioned works by taking into account both English and Chinese data to obtain a more comprehensive perception of the online discourse towards a given nation, namely, China. Moreover, data used in this research was event-centered. Collected during the significant event of the 70th anniversary of the P.R. China, online opinions are supposed to be more explicit and polarized, which would in turn benefit further analysis.

\section{Data Sets and Methods}
\label{sec:3}
This study is based on the Twitter data collected between 8 a.m. September 30 and 10 a.m. October 3. In this part, data acquisition and labeling are introduced in detail.
\subsection{Streaming API and Hashtags}
\label{sec:4}
The data sets used in this study were collected through the Twitter Streaming API. This kind of API allows to collect real-time tweets that incorporate specific hashtags, thus enables a comprehensive and targeted collection of tweets. In order to automatically capture the Chinese National Day-related tweets, over 1000 relevant tweets were studied and 59 frequently used Chinese National Day-related hashtags were extracted. These hashtags are shown in Table 1, and they are in English, Simplified Chinese, and Traditional Chinese.

\begin{table}
\caption{Hashtags used in this study}
\label{tab:1}       
\begin{tabular}{p{0.22\columnwidth}p{0.71\columnwidth}}  
\hline\noalign{\smallskip}
Language & Hashtags  \\
\noalign{\smallskip}\hline\noalign{\smallskip}
English & \#Beijing, \#CCP, \#CCP70, \#CCP70Bday, \#CelebrateChina70, \#China, \#china, \#Chinaat70, \#China70, \#China70years, \#Chinese, \#CPC, \#HiChina, \#MilitaryParade, \#NationalDay, \#NationalDay2019, \#NewChina70Years, \#PLA, \#PRC, \#prc, \#PRC70, \#prc70, \#PRC70thAnniv, \#PRC70YearsOn, \#ReformandOpeningUp, \#SeeChina, \#ThisIsChina, \#70thAnniversary, \#70YearsProsperity \\
Simplified Chinese & \#\begin{CJK}{UTF8}{gbsn} 北京 \end{CJK}, \#\begin{CJK}{UTF8}{gbsn} 大阅兵 \end{CJK} \#\begin{CJK}{UTF8}{gbsn} 国庆 \end{CJK}, \#\begin{CJK}{UTF8}{gbsn} 国庆阅兵 \end{CJK}, \#\begin{CJK}{UTF8}{gbsn} 国庆70年 \end{CJK}, \#\begin{CJK}{UTF8}{gbsn} 七十周年 \end{CJK}, \#\begin{CJK}{UTF8}{gbsn} 十一国庆 \end{CJK}, \#\begin{CJK}{UTF8}{gbsn} 中共 \end{CJK}, \#\begin{CJK}{UTF8}{gbsn} 中国 \end{CJK}, \#\begin{CJK}{UTF8}{gbsn} 中国国庆 \end{CJK}, \#\begin{CJK}{UTF8}{gbsn} 中华人民共和国 \end{CJK}, \#\begin{CJK}{UTF8}{gbsn} 中华人民共和国成立70周年 \end{CJK}, \#\begin{CJK}{UTF8}{gbsn} 阅兵 \end{CJK}, \#\begin{CJK}{UTF8}{gbsn} 祖国万岁 \end{CJK}, \#\begin{CJK}{UTF8}{gbsn} 70周年 \end{CJK} \\
Traditional Chinese & \#\begin{CJK}{UTF8}{bsmi} 北京 \end{CJK}, \#\begin{CJK}{UTF8}{bsmi} 大閱兵 \end{CJK}, \#\begin{CJK}{UTF8}{bsmi} 國慶 \end{CJK}, \#\begin{CJK}{UTF8}{bsmi} 國慶閱兵 \end{CJK}, \#\begin{CJK}{UTF8}{bsmi} 國慶70年 \end{CJK}, \#\begin{CJK}{UTF8}{bsmi} 七十週年 \end{CJK}, \#\begin{CJK}{UTF8}{bsmi} 十一國慶 \end{CJK}, \#\begin{CJK}{UTF8}{bsmi} 中共 \end{CJK}, \#\begin{CJK}{UTF8}{bsmi} 中國 \end{CJK}, \#\begin{CJK}{UTF8}{bsmi} 中國國慶 \end{CJK}, \#\begin{CJK}{UTF8}{bsmi} 中華人民共和國 \end{CJK}, \#\begin{CJK}{UTF8}{bsmi} 中華人民共和國成立70週年 \end{CJK}, \#\begin{CJK}{UTF8}{bsmi} 閱兵 \end{CJK}, \#\begin{CJK}{UTF8}{bsmi} 祖國萬歲 \end{CJK}, \#\begin{CJK}{UTF8}{bsmi} 70週年 \end{CJK},  \\
\noalign{\smallskip}\hline
\end{tabular}
\end{table}

The hashtags above were later applied to the Streaming API to retrieve the tweets containing them. After 74 hours of data collection, a total of 311,935 tweets, written in 54 languages, were obtained. English tweets accounted for 249,602 and Chinese (including simplified and traditional) tweets contributed to another 5,150. The remaining tweets are outside the scope of this article. Therefore, they are excluded from the following steps. Examples of collected tweets are shown in table 2.

\subsection{Data Labeling}
\label{sec:5}
Labeled data sets are necessary for supervised machine learning. Considering the large scale of the data, 1000 tweets were randomly sampled from either language, that is, 1000 English tweets and 1000 Chinese tweets. Then, four labelers with diverse backgrounds were involved in the labeling work. Detailed information about the labelers are shown in table 3. For every tweet within the 1000 English tweets, each labeler needed to independently decide whether it was relevant to the 2019 Chinese National Day, and if so, rate its sentiment towards China on a scale of -2 to +2. Positive number means a positive sentiment while negative number represents negative attitude. The absolute value of the score depicts the degree of the emotion. For example, a tweet labeled as +2 usually has extremely positive view towards China, and a tweet with a -1 shows negative opinion. Zero indicates that the tweet is objective and does not present obvious sentiment. For Chinese tweets, the procedure was the same except that the labelers needed to additionally judge whether the tweet was written in simplified or traditional Chinese.

\begin{table}
\caption{Examples for collected tweets (only show the time and text part)}
\label{tab:2}       
\renewcommand{\arraystretch}{1.5}
	\begin{tabular}{p{0.10\columnwidth}p{0.83\columnwidth}}
		\hline\noalign{\smallskip}
		Language & Tweets \\
		\noalign{\smallskip}\hline\noalign{\smallskip}
		\multirow{2}*{English} & "created\_at":"Tue Oct 01 16:58:48 +0000 2019" \\
		~ & "text":"@goofrider Big Congratulations to \#China70years" \\
		\cline{2-2}
		~ & "created\_at":"Tue Oct 01 11:00:43 +0000 2019" \\
		~ & "text":"Happy 70th Anniversary to China. Best wishes from Pakistan. \#China70years  \#China" \\
		\cline{2-2}
		~ & "created\_at":"Tue Oct 01 14:44:33 +0000 2019" \\
		~ & "text":"RT @HappsNews: NOW: Pro-democracy protests are happening all over \#HongKong on the 70th anniversary of \#China's \#NationalDay. " \\
		\noalign{\smallskip}\hline\noalign{\smallskip}
		\multirow{2}*{Chinese} & "created\_at":"Mon Sep 30 18:19:59 +0000 2019" \\
		~ & "text":"\#\begin{CJK}{UTF8}{gbsn} 中国 \end{CJK} \#\begin{CJK}{UTF8}{gbsn} 国庆节快乐 \end{CJK}\begin{CJK}{UTF8}{bsmi} 祝祖国母亲70周年生日快乐 \end{CJK} https://t.co/OXK0jZkA05" \\
		\cline{2-2}
		~ & "created\_at":"Tue Oct 01 06:00:13 +0000 2019" \\
		~ & "text":"\#China70years \begin{CJK}{UTF8}{gbsn} 想找到和祖国连接的点那就只有好好找到自己对于国家而言的价值了 \end{CJK}" \\
		\cline{2-2}
		~ & "created\_at":"Tue Oct 01 01:30:42 +0000 2019" \\
		~ & "text": "RT @RFA\_Chinese: \begin{CJK}{UTF8}{gbsn} 【国庆日，香港市面死寂】10月1日，政府制定多项临时措施，防止群众聚集。多个地铁站全日封闭，机场采取特别交通措施，多个市区大型商场停止营业，而街道上亦没有像往年国庆一样挂满国旗。有市民表示，国庆日，香港跟戒严没有分别。 \end{CJK}\#\begin{CJK}{UTF8}{gbsn} 十一国庆 \end{CJK}\#\begin{CJK}{UTF8}{gbsn} 戒严 \end{CJK}" \\
		\hline
	\end{tabular}
\end{table}

\begin{table}
\caption{The backgrounds of the labeler}
\label{tab:3}       
\begin{tabular}{llll}
\hline\noalign{\smallskip}
Labeler & Comes from & Major & Sex \\
\noalign{\smallskip}\hline\noalign{\smallskip}
1 & Mainland China & Data Science & Male \\
2 & Taiwan & Environmental Engineering & Female \\
3 & Mainland China & Accounting & Female \\
4 & Mainland China & Law & Male \\
\noalign{\smallskip}\hline
\end{tabular}
\end{table}

\begin{table}
\centering
\caption{The voting ratio of 4 labelers} 
\label{tab:4}       
\renewcommand{\arraystretch}{1.5}
\begin{tabular}{|p{0.08\columnwidth}|p{0.12\columnwidth}|p{0.12\columnwidth}|p{0.12\columnwidth}|p{0.12\columnwidth}|p{0.06\columnwidth}|p{0.06\columnwidth}|p{0.12\columnwidth}|} 
\hline 
\begin{center} Language of tweets \end{center} & \multicolumn{2}{p{0.24\columnwidth}|}{\begin{center}Decisions on\end{center}} & \begin{center} 4:0 (3:0 for row 3, 5, 8 and 10) \end{center} & \begin{center} 3:1 (2:1 for row 3, 5, 8 and 10) \end{center} & \begin{center} 2:2 \end{center} & \begin{center} 2:1:1 \end{center} & \begin{center} 1:1:1:1 (1:1:1 for row 3, 5, 8 and 10) \end{center} \\ 
\hline 
\begin{center} \multirow{5}*{English} \end{center} & \multicolumn{2}{p{0.24\columnwidth}|}{\begin{center}Relevance \end{center}} & \begin{center} 880 \end{center} & \begin{center} 92 \end{center} & \begin{center} 28 \end{center} & \begin{center} 0 \end{center} & \begin{center} 0 \end{center} \\
\cline{2-8}
~ & \begin{center} \multirow{2}*{\shortstack{Sentiment (on a\\-2 to +2 scale)}} \end{center} & \begin{center} If the voting result of relevance is 4:0 \end{center} & \begin{center} 53 \end{center} & \begin{center} 164 \end{center} & \begin{center} 64 \end{center} & \begin{center} 31 \end{center} & \begin{center} 1 \end{center} \\
\cline{3-8}
~ & ~ & \begin{center} If the voting result of relevance is 3:1 \end{center} & \begin{center} 12 \end{center} & \begin{center} 53 \end{center} & \begin{center}/ \end{center} & \begin{center} / \end{center} & \begin{center} 6 \end{center} \\
\cline{2-8}
~ & \begin{center} \multirow{2}*{\shortstack{Sentiment (in\\the positive,\\ neutral, negative\\scheme)}} \end{center} & \begin{center} If the voting result of relevance is 4:0 \end{center} & \begin{center} 202 \end{center} & \begin{center} 81 \end{center} & \begin{center} 23 \end{center} & \begin{center} 7 \end{center} & \begin{center} 0 \end{center} \\
\cline{3-8}
~ & ~ & \begin{center} If the voting result of relevance is 3:1 \end{center} & \begin{center} 41 \end{center} & \begin{center} 29 \end{center} & \begin{center} / \end{center} & \begin{center} / \end{center} & \begin{center} 1 \end{center} \\
\hline
\begin{center} \multirow{6}*{Chinese} \end{center} & \multicolumn{2}{p{0.24\columnwidth}|}{\begin{center}Relevance \end{center}} & \begin{center} 820 \end{center} & \begin{center} 120 \end{center} & \begin{center} 59 \end{center} & \begin{center} 0 \end{center} & \begin{center} 0 \end{center} \\
\cline{2-8}
~ & \begin{center} \multirow{2}*{\shortstack{Sentiment (on a\\-2 to +2 scale)}} \end{center} & \begin{center} If the voting result of relevance is 4:0 \end{center} & \begin{center} 98 \end{center} & \begin{center} 159 \end{center} & \begin{center} 82 \end{center} & \begin{center} 37 \end{center} & \begin{center} 3 \end{center} \\
\cline{3-8}
~ & ~ & \begin{center} If the voting result of relevance is 3:1 \end{center} & \begin{center} 17 \end{center} & \begin{center} 51 \end{center} & \begin{center} / \end{center} & \begin{center} / \end{center} & \begin{center} 11 \end{center} \\
\cline{2-8}
~ & \begin{center} \multirow{2}*{\shortstack{Sentiment (in\\the positive,\\ neutral, negative\\scheme)}} \end{center} & \begin{center}If the voting result of relevance is 4:0 \end{center} & \begin{center} 309 \end{center} & \begin{center} 48 \end{center} & \begin{center} 17 \end{center} & \begin{center} 5 \end{center}  & \begin{center} 0 \end{center} \\
\cline{3-8}
~ & ~ & \begin{center} If the voting result of relevance is 3:1 \end{center} & \begin{center} 51 \end{center} & \begin{center} 24 \end{center} & \begin{center} / \end{center} & \begin{center} / \end{center} & \begin{center} 4 \end{center}\\
\cline{2-8}
~ & \multicolumn{2}{p{0.24\columnwidth}|}{\begin{center} Simplified or traditional \end{center}} & \begin{center} 942 \end{center} & \begin{center} 43 \end{center} & \begin{center} 14 \end{center} & \begin{center} 0 \end{center} & \begin{center} 0 \end{center} \\
\hline 
\end{tabular}
\end{table}

\begin{table}
\centering
\caption{Comparison between decisions made by labeler 2 and the overall voting results} 
\label{tab:5}       
\renewcommand{\arraystretch}{1.5}
\begin{tabular}{|p{0.10\columnwidth}|p{0.07\columnwidth}|p{0.12\columnwidth}|p{0.12\columnwidth}|p{0.07\columnwidth}|p{0.12\columnwidth}|p{0.12\columnwidth}|p{0.08\columnwidth}|} 
\hline 
\begin{center} Language of tweets \end{center} & \multicolumn{3}{p{0.36\columnwidth}|}{\begin{center}English\end{center}} & \multicolumn{4}{p{0.36\columnwidth}|}{\begin{center}Chinese\end{center}} \\ 
\hline
\begin{center} \multirow{2}*{Decisions on} \end{center} & \begin{center} \multirow{2}*{Relecance} \end{center} & \multicolumn{2}{p{0.24\columnwidth}|}{\begin{center}\shortstack{Sentiment (in the positive,\\ neutral, negative scheme)} \end{center}} & \begin{center} \multirow{2}*{Relecance} \end{center} & \multicolumn{2}{p{0.12\columnwidth}|}{\begin{center}\shortstack{Sentiment (in the positive,\\ neutral, negative scheme)} \end{center}} & \begin{center} \multirow{2}*{\shortstack{Simplified\\or\\traditional}} \end{center} \\
\cline{3-4}
\cline{6-7}
~ & ~ & \begin{center} If the voting result of relevance is 4:0 \end{center} & \begin{center} If the voting result of relevance is 3:1 \end{center} & ~ & \begin{center} If the voting result of relevance is 4:0 \end{center} & \begin{center} If the voting result of relevance is 3:1 \end{center} & ~ \\
\hline
\begin{center} Number of votes that the result is 3:1 or 2:1 * \end{center} & \begin{center} 92 \end{center}  & \begin{center}81 \end{center}  & \begin{center} 29 \end{center}  & \begin{center} 120 \end{center}  & \begin{center} 48 \end{center}  & \begin{center} 19 \end{center}  & \begin{center} 43 \end{center} \\
\hline
\begin{center} Number of cases that labeler 2 had different opinion with the other three or two labelers \end{center} & \begin{center} 6 \end{center}  & \begin{center}20 \end{center}  & \begin{center} 11 \end{center}  & \begin{center} 18 \end{center}  & \begin{center} 23 \end{center}  & \begin{center} 11 \end{center}  & \begin{center} 4 \end{center} \\
\hline
\begin{center} Ratio of labeler 2 being the dissident \end{center} & \begin{center} 6.5\% \end{center}  & \begin{center}24.7\% \end{center}  & \begin{center} 37.9\% \end{center}  & \begin{center} 15.0\% \end{center}  & \begin{center} 47.9\% \end{center}  & \begin{center} 57.9\% \end{center}  & \begin{center} 9.3\% \end{center} \\
\hline
\begin{center} Overall ratio \end{center} & \multicolumn{7}{p{0.80\columnwidth}|}{\begin{center}21.5\% \end{center}} \\
\hline
\multicolumn{8}{|p{0.95\columnwidth}|}{* If labeler 2 voted for irrelevance in the first place and thus did not have a vote in the sentiment part, it would not be counted in the number shown in column 4 and 7 due to the absence of labeler 2. In other words, this table only focuses on the voting that labeler 2 has participated. } \\
\hline
\end{tabular}
\end{table}

The results of four labelers were then amalgamated through voting. Before merging the data sets, the voting ratio were counted. The results are displayed in Table 4. From table 4, it can be found that in most cases for relevance and the types of Chinese characters (simplified or simplified), four labelers tend to be consistent. Only in less than 0.6\% of all cases did the voting results of the labelers show a tie (2:2). However, when it comes to the voting of sentiments, the labelers showed greater disagreement and made different decisions on most tweets. Thus, in order to make the labeling result more reliable and concrete, which is critical for the following supervised learning, the final label employs a positive-neutral-negative scheme. As is depicted in row 5, 6, 10 and 11 of table 4, the disagreement between four labelers was significantly narrowed after adopting the positive-neutral-negative scheme. 

Another issue that needs to be addressed is whether anyone of the four labelers typically holds different opinions with the other three. That is, would the answers provided by a specific labeler always fail in the voting. Due to the balance of genders and majors, the places that the labelers came from constitute the main concern. Therefore, the decisions of labeler 2 were compared with the overall voting results and the outcomes are shown in table 5. As can be found in the table, the overall ratio that labeler 2 making different choices with others is 21.5\%, which does not indicate significant level of being isolated. Although labeler 2 tended to share different opinions when it comes to Chinese tweets, considering the fact that for 78.6\% of Chinese tweets all the four labelers made the same decisions on sentiments and this table only shows the 3:1 or 2:1 cases which accounts for merely 15.7\%, the influence of this difference could be minimal.

In most cases, since the decisions made by four labelers converge, the final label was determined without controversy. When faced with rare situations of 2:2 or 1:1:1:1, the final label would be chosen randomly from the two or four. 

\subsection{Methods}
\label{sec:6}
The technique used to analyze the general sentiments of the tweets is a mixed method of dictionary and SVM, which is an improved version of Tannier’s classifiers [22] and yields satisfactory results while avoid being too complicated. All the classifiers mentioned below are combinations of a trained SVM and positive-negative dictionaries.

Firstly, a relevance classifier would decide whether a tweet is relevant to the topic. For example, some tweets may be talking about the French National Day or the recent disputes and chaos in Hong Kong and were mistakenly fetched during the data collection step. These irrelevant tweets shall be excluded in the following analysis. Secondly, tweets written in Chinese would be categorized according to their typeface, that is, simplified or traditional Chinese. Thirdly, another classifier would judge whether the tweet is simply stating a fact or is expressing certain kinds of affection. Those tweets with affection would then be sent to the last classifier which determines the sentiment, positive or negative, of the tweet. 

\section{Experiments and Results}

\subsection{Performances of the Classifiers}
\label{sec:7}
After training the SVM on the labeled data set, as well as creating and adjusting the dictionaries manually, all these classifiers have reached an average accuracy of over 93\%. The detailed performances of the classifiers are shown in table 6.  

\begin{table}
\centering
\caption{The performances of the classifiers} 
\label{tab:6}       
\renewcommand{\arraystretch}{1.5}
\begin{tabular}{|p{0.22\columnwidth}|p{0.22\columnwidth}|p{0.22\columnwidth}|p{0.22\columnwidth}|} 
\hline 


\multicolumn{4}{|p{0.96\columnwidth}|}{\textbf{Relevance Classifier (for English Tweets)}} \\
\hline
\textbf{Label} & \textbf{Precision} & \textbf{Recall} & \textbf{F1} \\
\hline
Relevant & 0.943 & 0.984 & 0.963 \\
\hline
Irrelevant & 0.984 & 0.941 & 0.962 \\
\hline
Accuracy & \multicolumn{3}{|p{0.72\columnwidth}|}{0.962} \\
\hline

\multicolumn{4}{|p{0.96\columnwidth}|}{\textbf{Classifier to Determine Whether the Text Has Emotions (for English Tweets)}} \\
\hline
\textbf{Label} & \textbf{Precision} & \textbf{Recall} & \textbf{F1} \\
\hline
True & 0.909 & 1.0 & 0.952 \\
\hline
False & 1.0 & 0.900 & 0.947 \\
\hline
Accuracy & \multicolumn{3}{|p{0.72\columnwidth}|}{0.950} \\
\hline

\multicolumn{4}{|p{0.96\columnwidth}|}{\textbf{Positive or Negative Sentiment Classifier (for English Tweets)}} \\
\hline
\textbf{Label} & \textbf{Precision} & \textbf{Recall} & \textbf{F1} \\
\hline
Positive & 0.974 & 1.0 & 0.987 \\
\hline
Negative & 1.0 & 0.973 & 0.986 \\
\hline
Accuracy & \multicolumn{3}{|p{0.72\columnwidth}|}{0.987} \\
\hline

\multicolumn{4}{|p{0.96\columnwidth}|}{\textbf{Relevance Classifier (for Chinese Tweets)}} \\
\hline
\textbf{Label} & \textbf{Precision} & \textbf{Recall} & \textbf{F1} \\
\hline
Relevant & 0.943 & 0.989 & 0.965 \\
\hline
Irrelevant & 0.988 & 0.941 & 0.964 \\
\hline
Accuracy & \multicolumn{3}{|p{0.72\columnwidth}|}{0.965} \\
\hline

\multicolumn{4}{|p{0.96\columnwidth}|}{\textbf{Classifier to Determine Whether the Text Has Emotions (for Chinese Tweets)}} \\
\hline
\textbf{Label} & \textbf{Precision} & \textbf{Recall} & \textbf{F1} \\
\hline
True & 0.889 & 1.0 & 0.941 \\
\hline
False & 1.0 & 0.875 & 0.933 \\
\hline
Accuracy & \multicolumn{3}{|p{0.72\columnwidth}|}{0.938} \\
\hline

\multicolumn{4}{|p{0.96\columnwidth}|}{\textbf{Positive or Negative Sentiment Classifier (for Chinese Tweets)}} \\
\hline
\textbf{Label} & \textbf{Precision} & \textbf{Recall} & \textbf{F1} \\
\hline
Positive & 0.954 & 0.985 & 0.969 \\
\hline
Negative & 0.984 & 0.953 & 0.968 \\
\hline
Accuracy & \multicolumn{3}{|p{0.72\columnwidth}|}{0.969} \\
\hline

\multicolumn{4}{|p{0.96\columnwidth}|}{\textbf{Typeface Classifier (for Chinese Tweets)}} \\
\hline
\textbf{Label} & \textbf{Precision} & \textbf{Recall} & \textbf{F1} \\
\hline
Simplified & 0.984 & 1.0 & 0.992 \\
\hline
Traditional & 1.0 & 0.983 & 0.991 \\
\hline
Accuracy & \multicolumn{3}{|p{0.72\columnwidth}|}{0.992} \\
\hline

\end{tabular}
\end{table}

After confirmation of the performances, these classifiers were then applied to the temporal and spatial analysis of all the English and Chinese tweets collected.

\subsection{Temporal Fluctuations of the Sentiments}
\label{sec:8}
In this part, the overall sentiments of the Chinese National Day-related tweets would be analyzed with regard to time. The abscissa in the figure is labeled with Beijing time.

Figure 1 and 2 shows the number of English and Chinese tweets concerning the Chinese National Day. The highest peak of English tweets was around 18 o’clock October 1 and the second peak was around 11 o’clock. The two most significant peak for Chinese tweets was around 10 o’clock and 20 o’clock October 1.

Figure 3 and 4 depicts the absolute sentiments of the English and Chinese tweets towards China. The absolute sentiments are calculated by subtracting the number of negative tweets from the number of positive tweets, therefore, a score above 0 indicates an overall positive sentiment towards China at the time period, and vice versa. As is shown in the pictures, the sentiments of the two languages are disparate.

Figure 5 and Figure 6 are the variation of normalized sentiments towards China with time. The blue line and the green line indicate the percentages of positive and negative tweets, respectively. The red line is the difference between the two, which represents the overall sentiment. The jitter of shown in Figure 6 may be due to the relatively small amount of data of Chinese tweets. In general, Chinese tweets are more positive than English tweets.

\subsection{Spatial Analysis of the Sentiments}
\label{sec:9}
In this section, the overall sentiments of the tweets would be studied based on the self-proclaimed location of the Twitter users. Every tweet returned by the Twitter API attached the corresponding user profile, which includes the location that the user claimed to be. Although this self-proclaimed location might not be the actual place the user lived, the global sentiment maps based on these locations still have some reference value. Also, to reduce the inaccuracy brought by this self-proclaim-error, the following analysis was conducted on the national level rather than local level.

Due to the large scale of the English tweets, it would be time-consuming to manually link all the locations written in the user profile with its nation. Therefore, 2000 tweets were randomly picked to be examined and labeled. The pictures shown in this section has proved that 2000 is enough for yielding meaningful results. The scale of Chinese tweets was much smaller than English tweets, and all of them were manually labeled.

Figure 7 to 10 shows the exact number of the positive or negative tweets posted by each country. Only countries with at least one tweet are marked. Figure 11 to 12 shows the comparisons of the numbers of positive and negative tweets. The score for each country was calculated by dividing number of positive tweets by the number of negative tweets. 

\subsection{Frequently Used Words by Different Sentiments}
\label{sec:10}
To provide an intuitive perception of the discourse expressed in the collected tweets, the frequencies of words for both positive and negative tweets were calculated. After removing the stop words, the histograms for the top 10 frequently used words for both languages and both sentiments are shown in figure 13 and 15. Also, the word clouds are provided in figure 14 and 16 after removing words shared by both parties.

\begin{figure}
  \includegraphics[width=16cm,height=6cm]{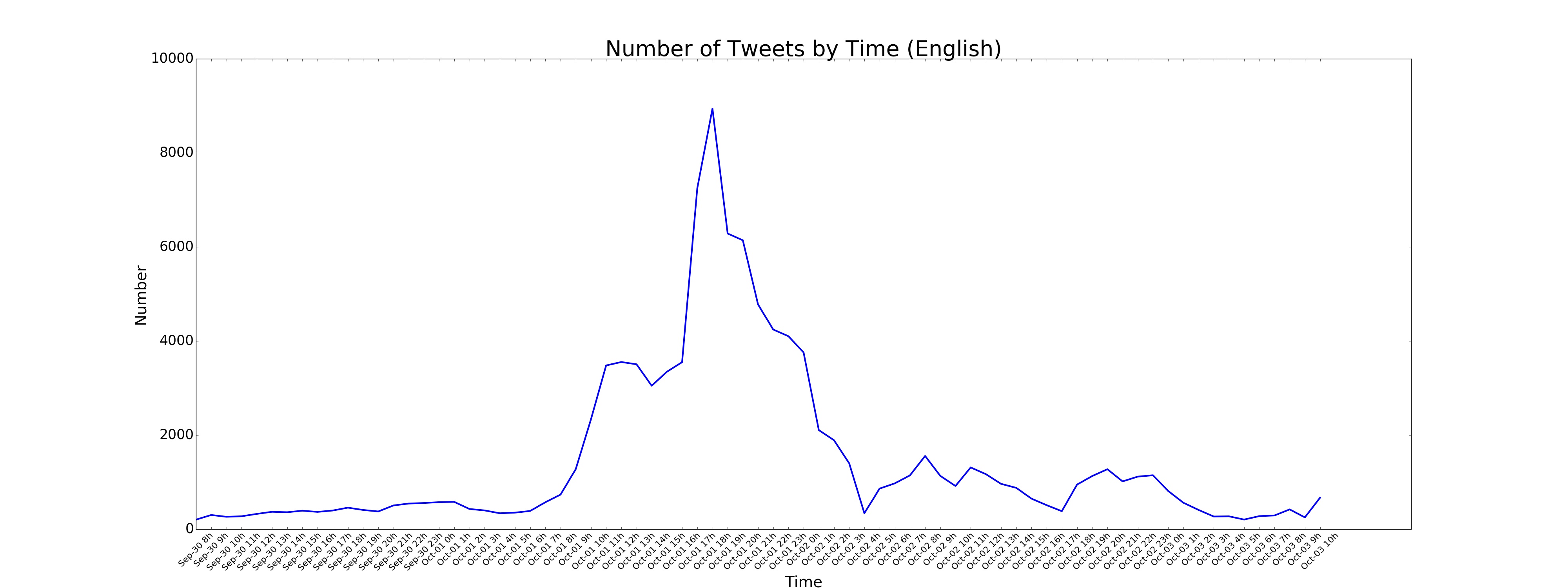}
\caption{Number of English Tweets by Time}
\label{fig:1}       
\end{figure}

\begin{figure}
  \includegraphics[width=16cm,height=6cm]{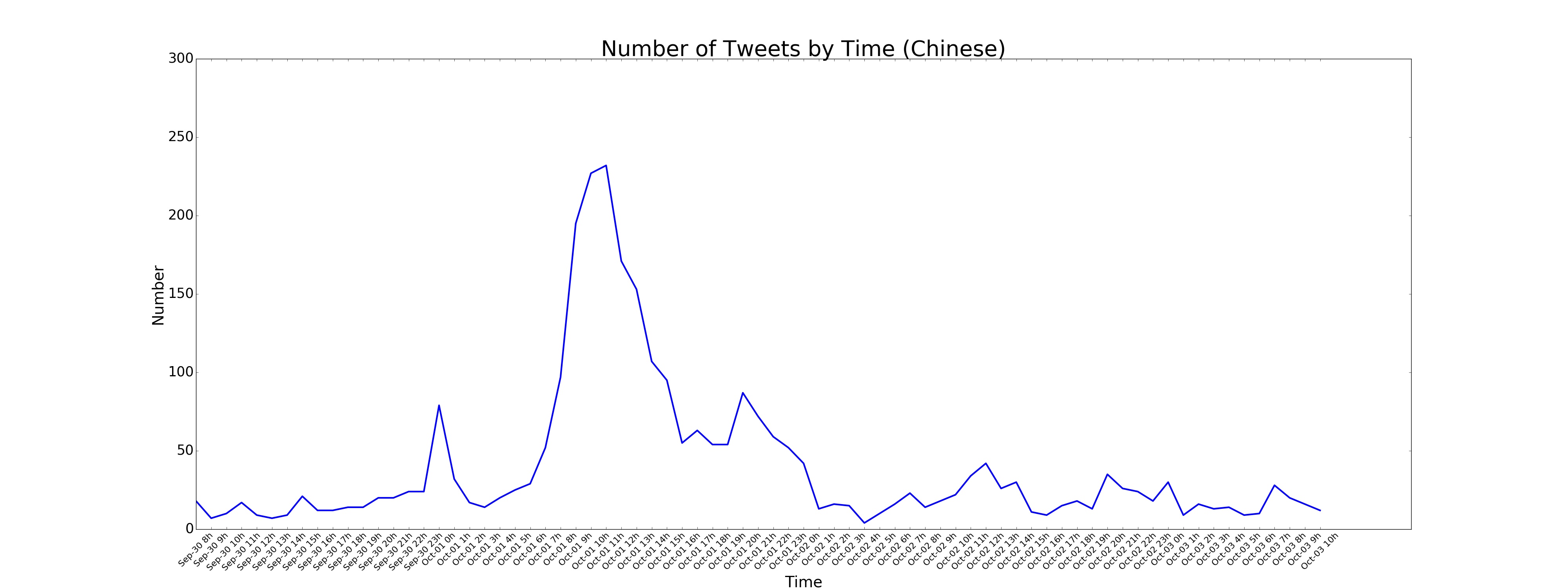}
\caption{Number of Chinese Tweets by Time}
\label{fig:2}       
\end{figure}

\begin{figure}
  \includegraphics[width=16cm,height=6cm]{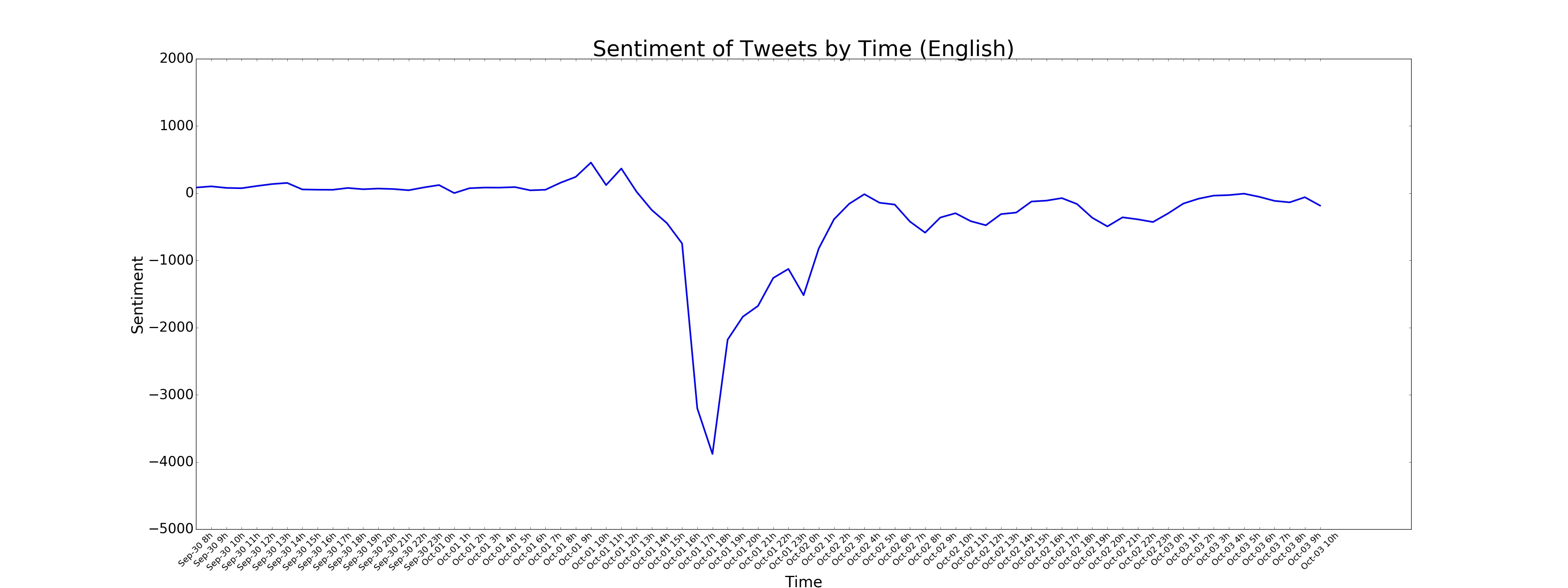}
\caption{Absolute Sentiments of English Tweets by Time}
\label{fig:3}       
\end{figure}

\begin{figure}
  \includegraphics[width=16cm,height=6cm]{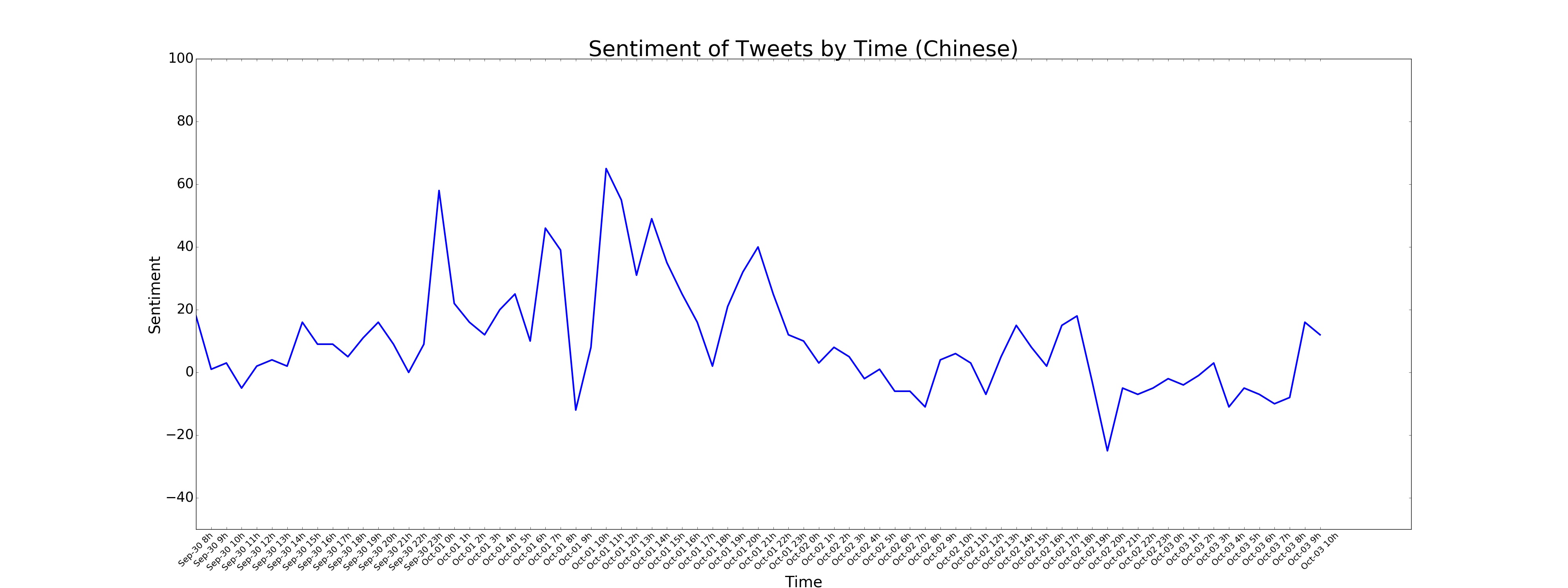}
\caption{Absolute Sentiments of Chinese Tweets by Time}
\label{fig:4}       
\end{figure}

\begin{figure}
  \includegraphics[width=16cm,height=6cm]{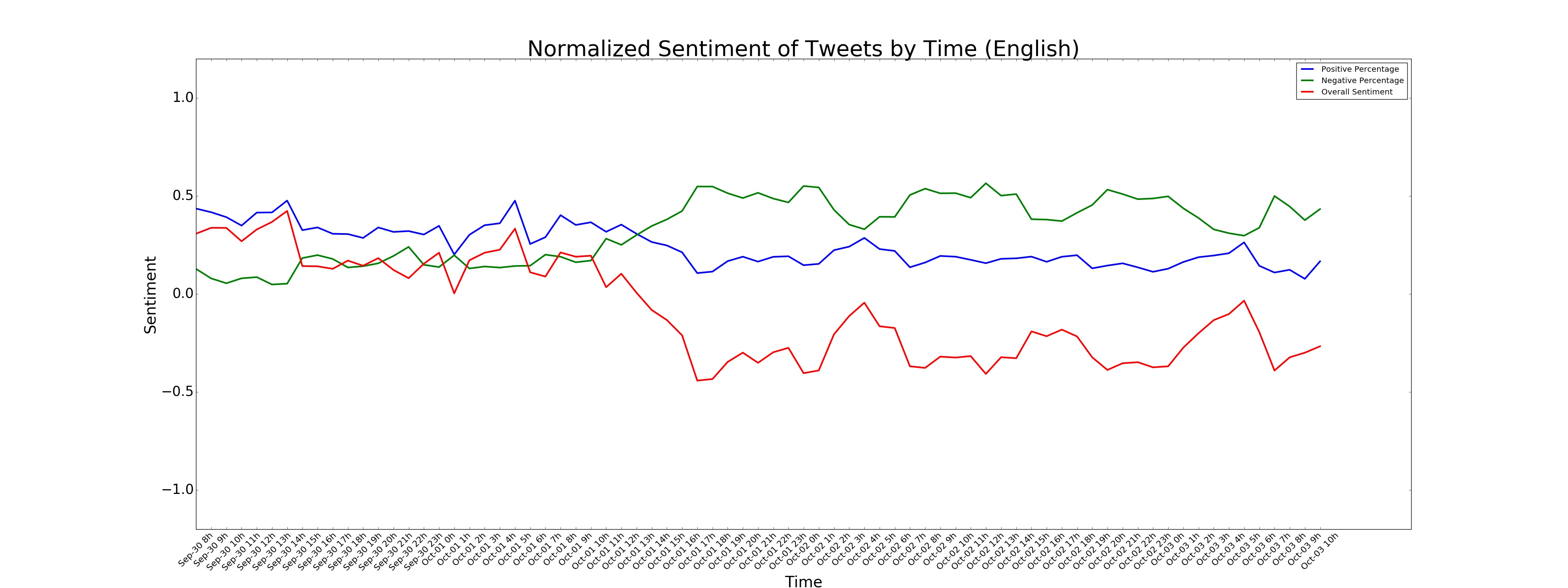}
\caption{Normalized Sentiments of English Tweets by Time}
\label{fig:5}       
\end{figure}

\begin{figure}
  \includegraphics[width=16cm,height=6cm]{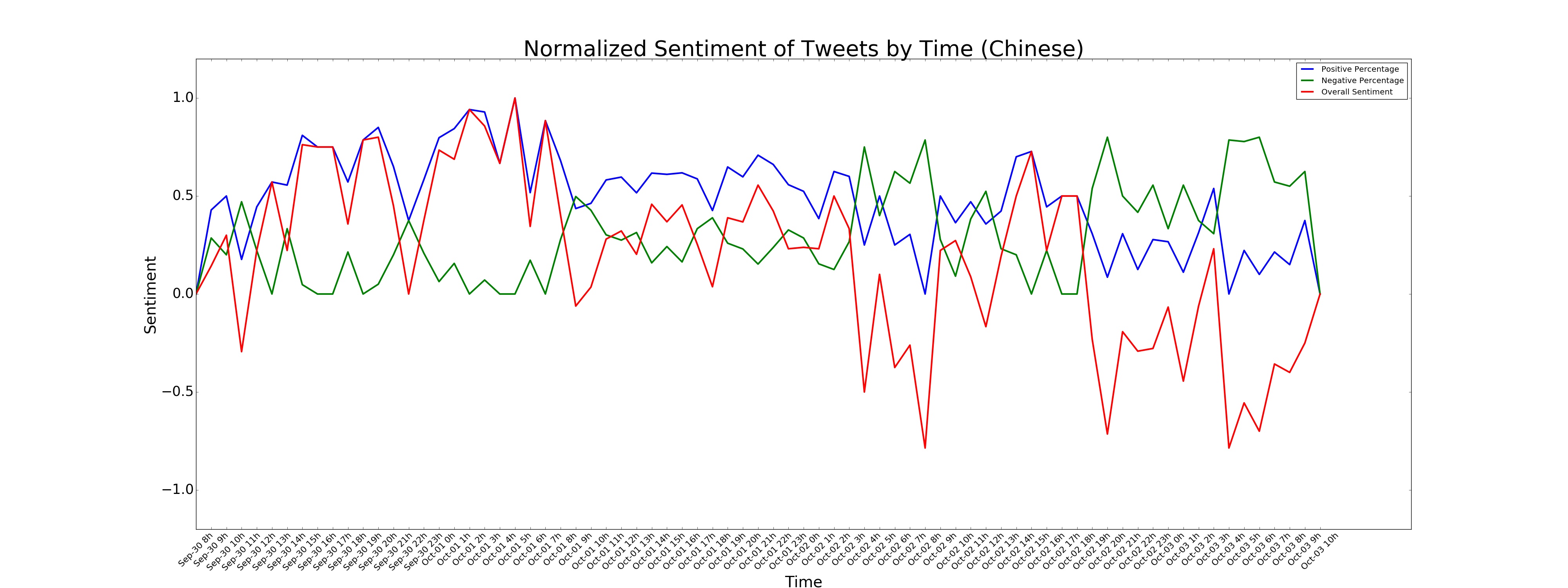}
\caption{Normalized Sentiments of Chinese Tweets by Time}
\label{fig:6}       
\end{figure}

\begin{figure}
  \includegraphics[width=12.1cm,height=6.6cm]{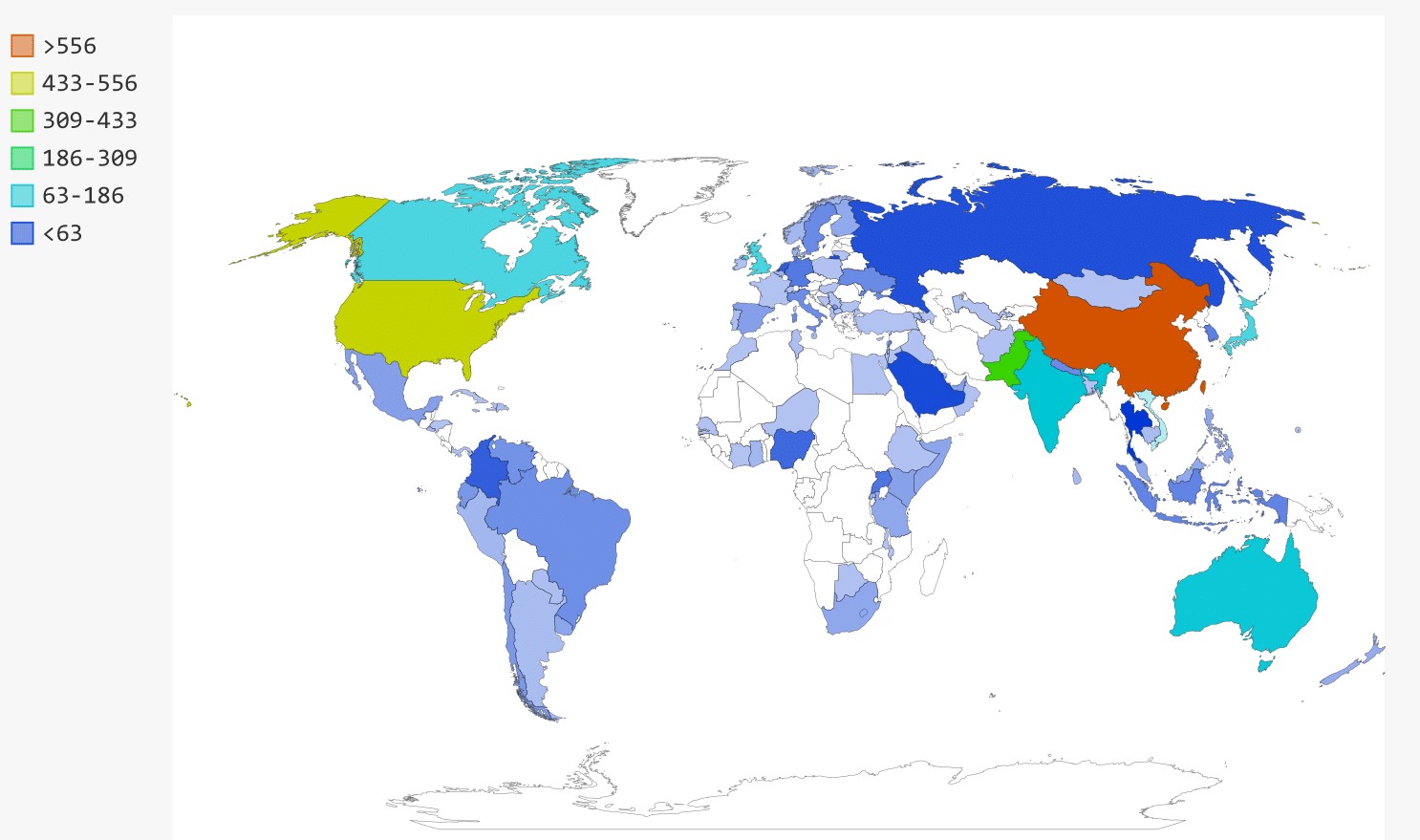}
\caption{The number of English positive tweets produced by each country}
\label{fig:7}       
\end{figure}

\begin{figure}
  \includegraphics[width=12.1cm,height=6.6cm]{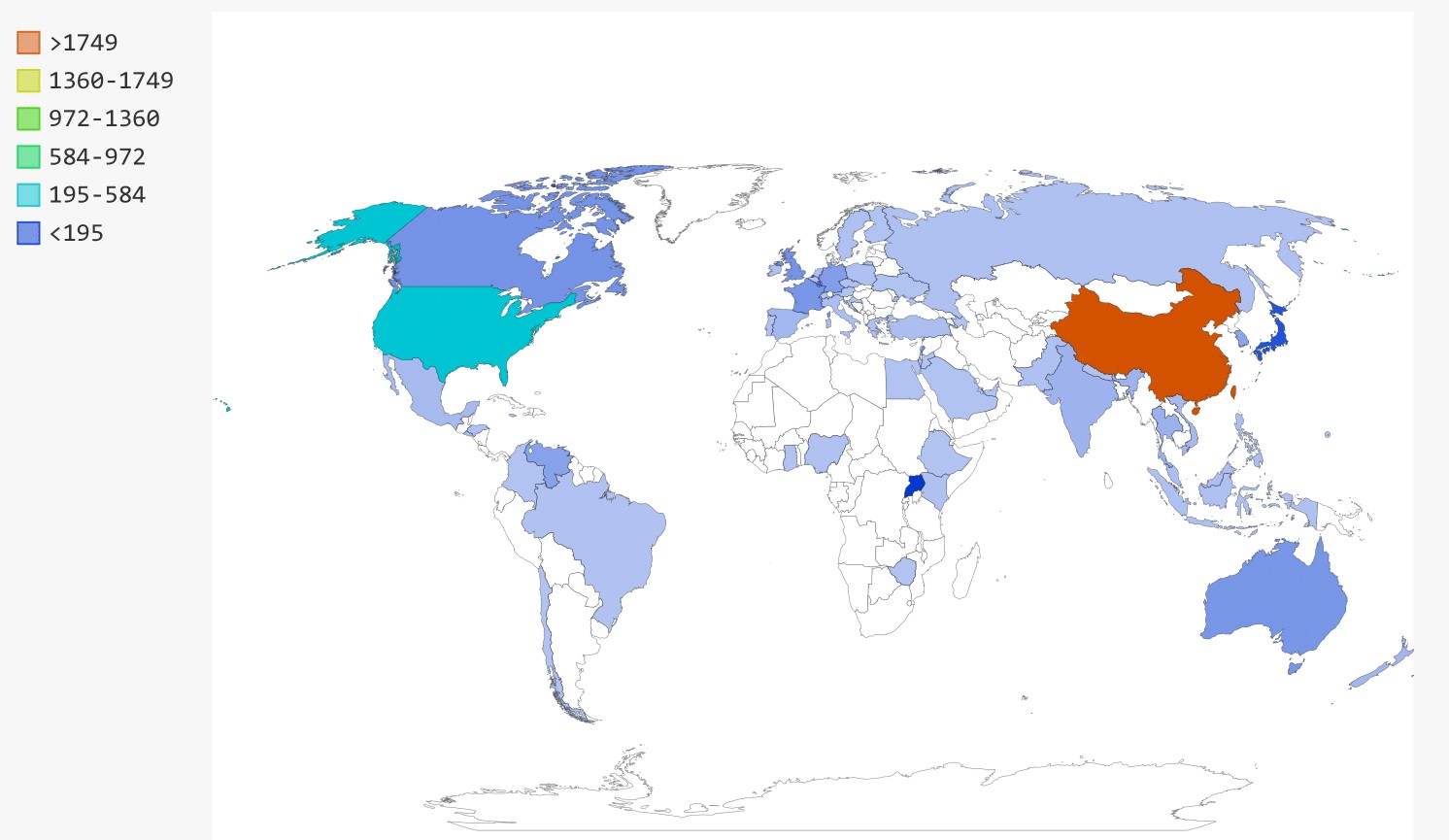}
\caption{The number of English negative tweets produced by each country}
\label{fig:8}       
\end{figure}

\begin{figure}
  \includegraphics[width=12.1cm,height=6.6cm]{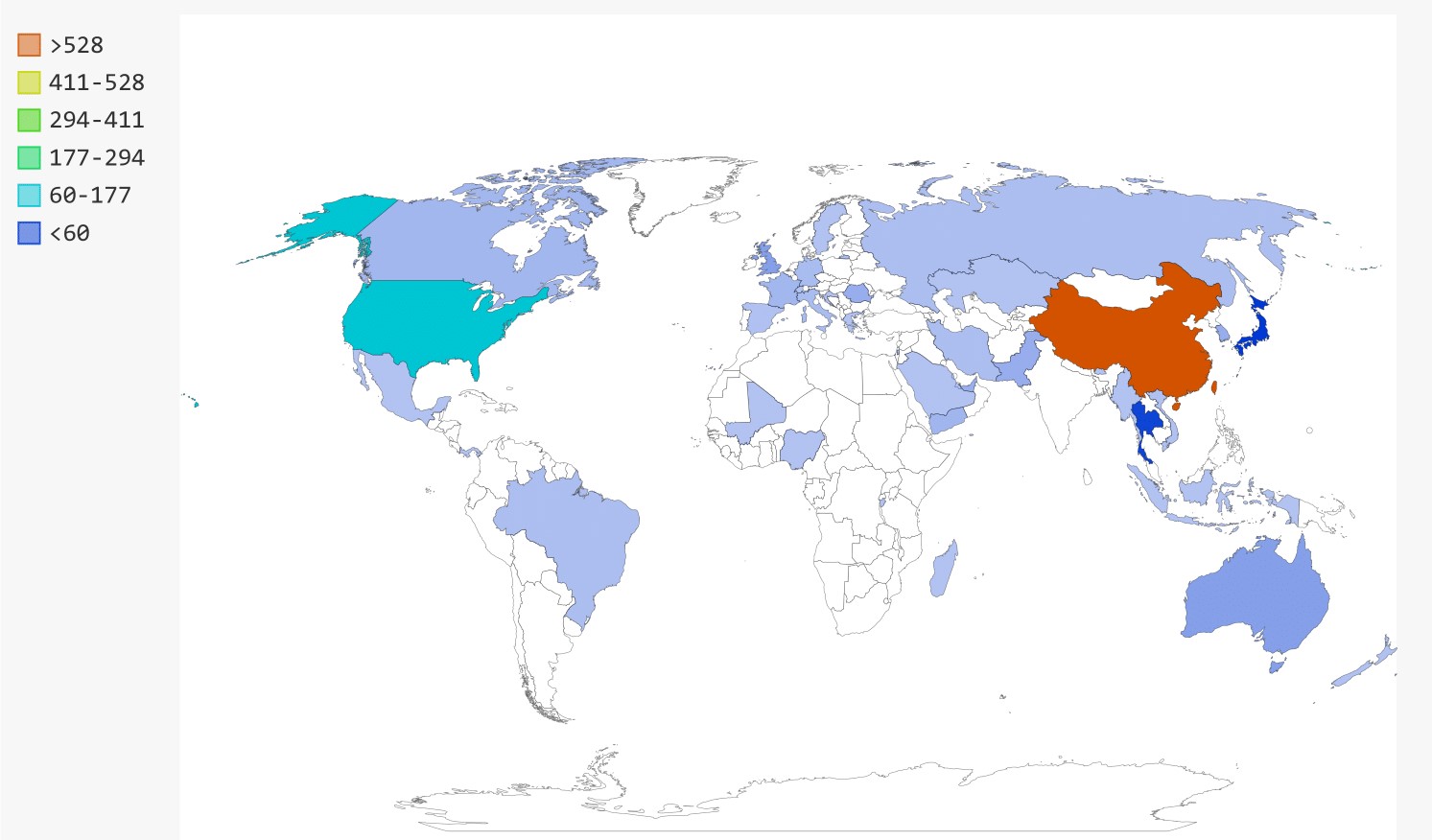}
\caption{The number of Chinese positive tweets produced by each country}
\label{fig:9}       
\end{figure}

\begin{figure}
  \includegraphics[width=12.1cm,height=6.6cm]{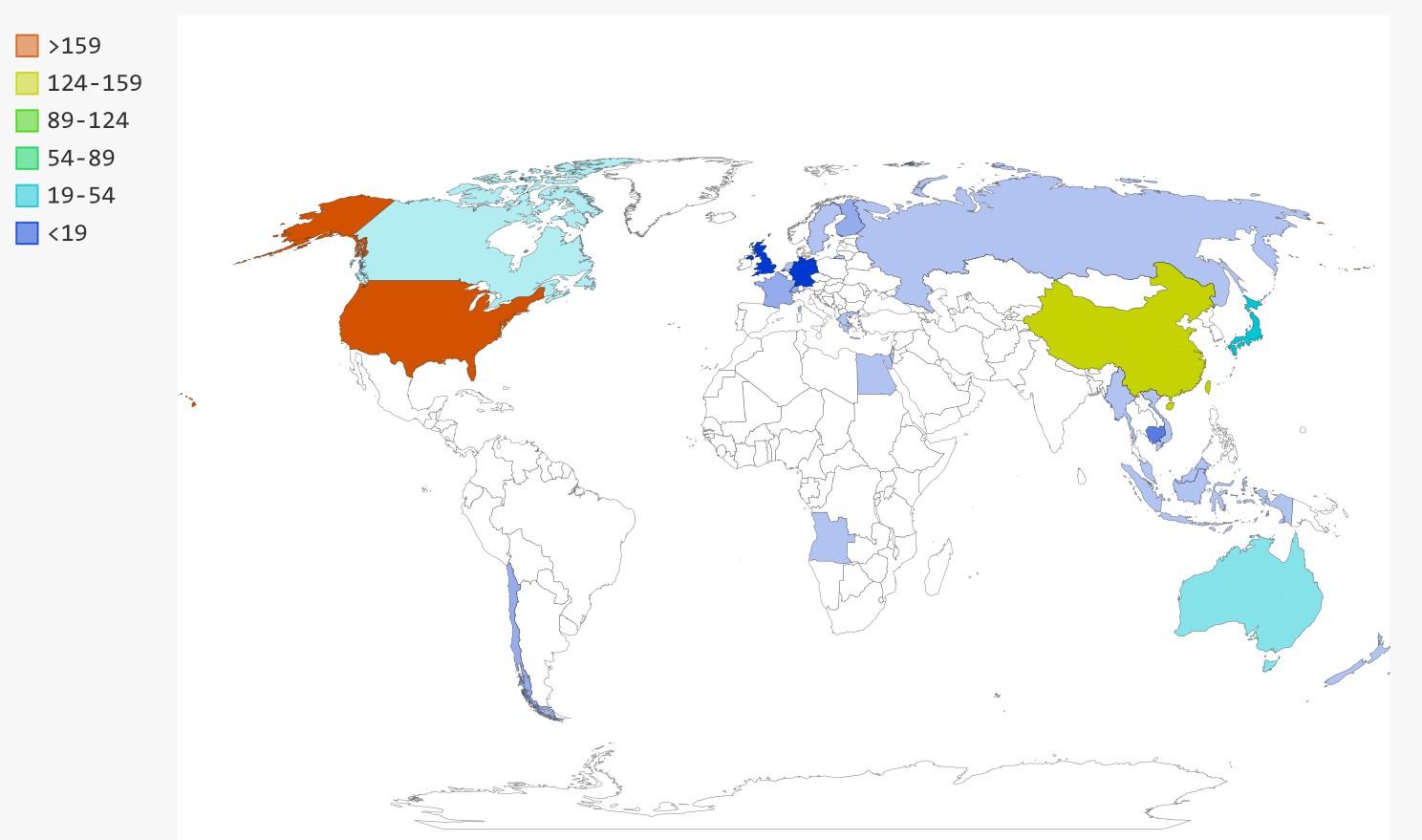}
\caption{The number of Chinese negative tweets produced by each country}
\label{fig:10}       
\end{figure}

\begin{figure}
  \includegraphics[width=12.1cm,height=6.6cm]{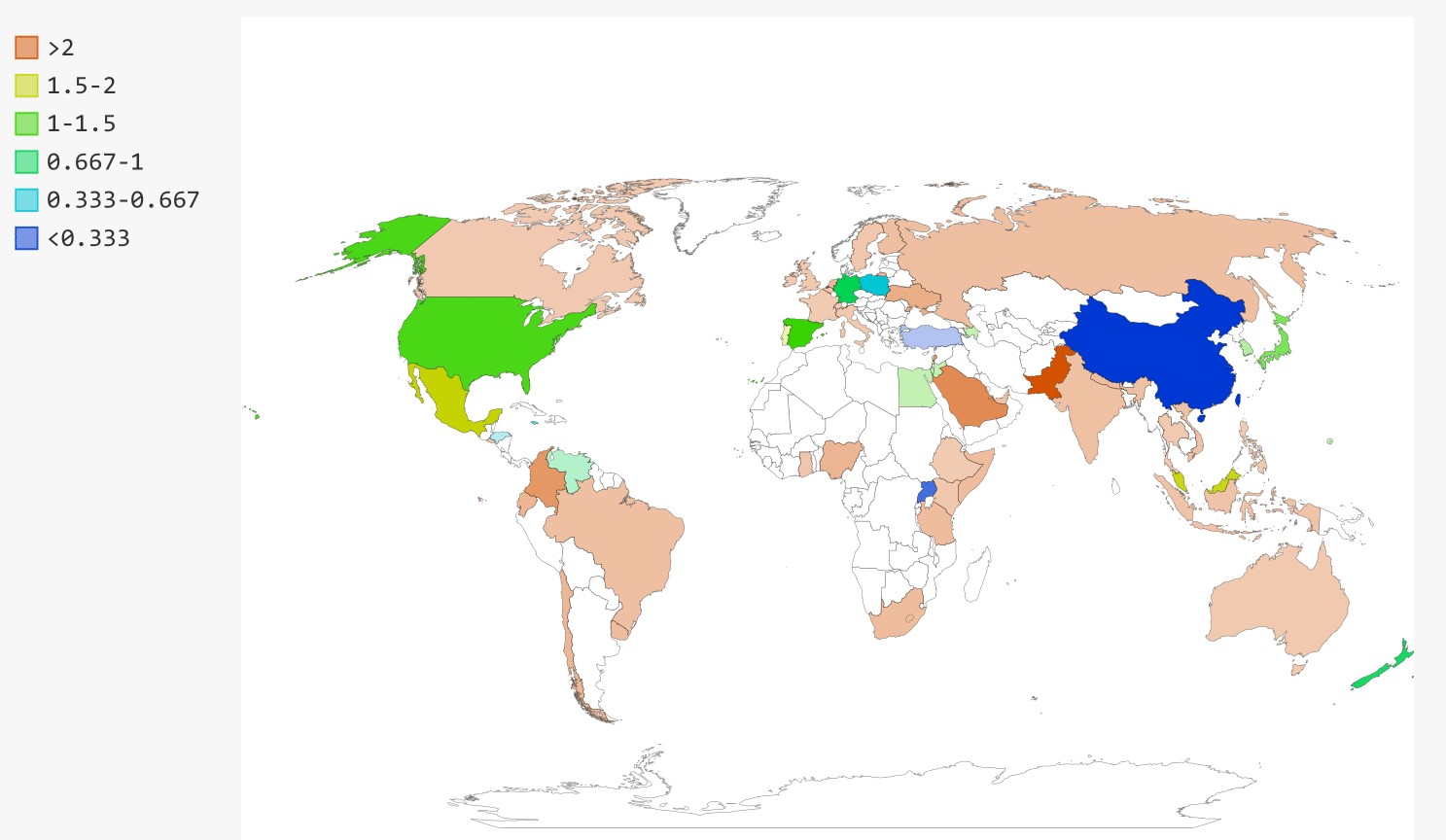}
\caption{The comparison of positive and negative English tweets produced by each country}
\label{fig:11}       
\end{figure}

\begin{figure}
  \includegraphics[width=12.1cm,height=6.6cm]{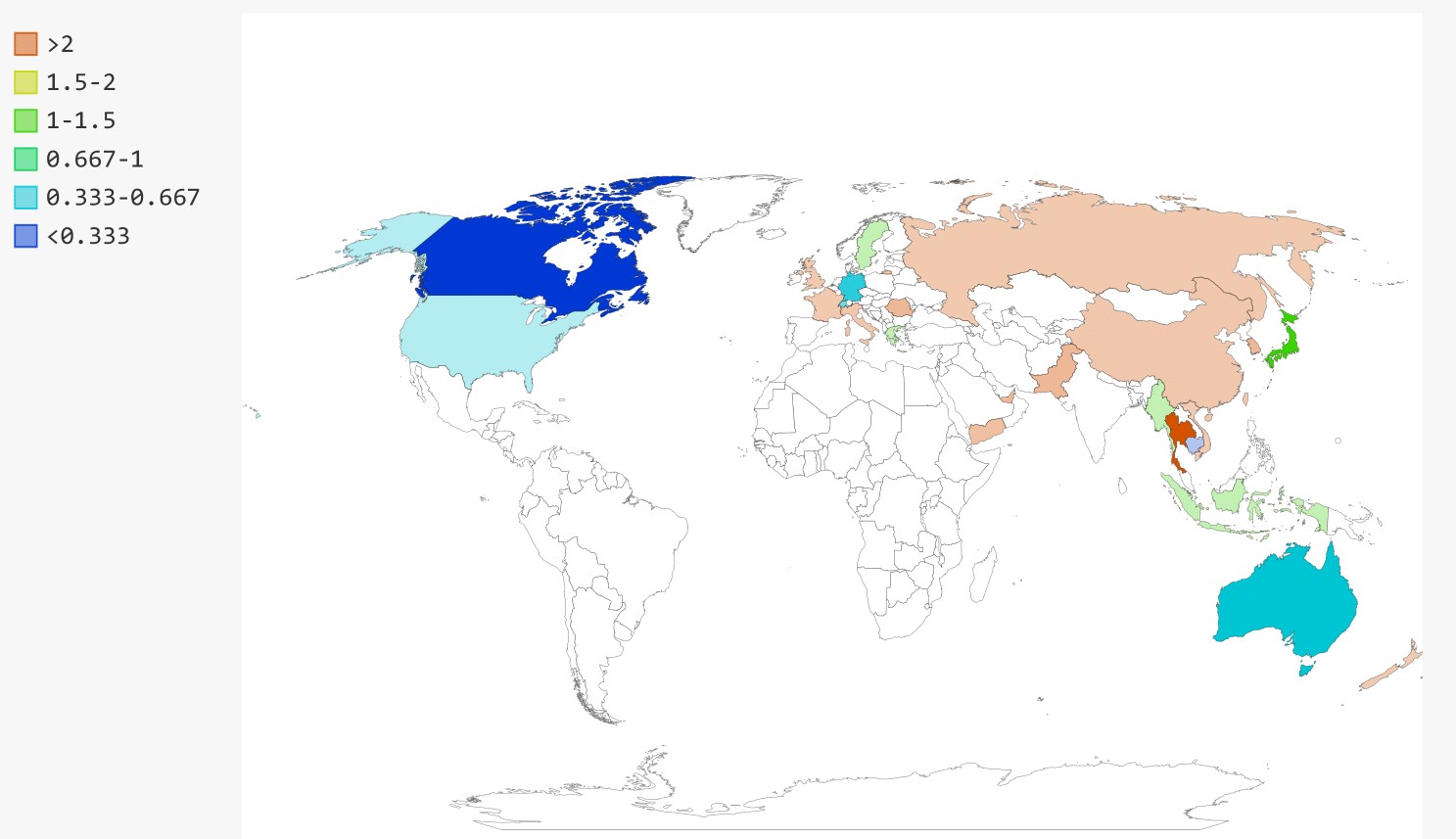}
\caption{The comparison of positive and negative Chinese tweets produced by each country}
\label{fig:12}       
\end{figure}

\begin{figure}
  \includegraphics[width=16cm,height=5cm]{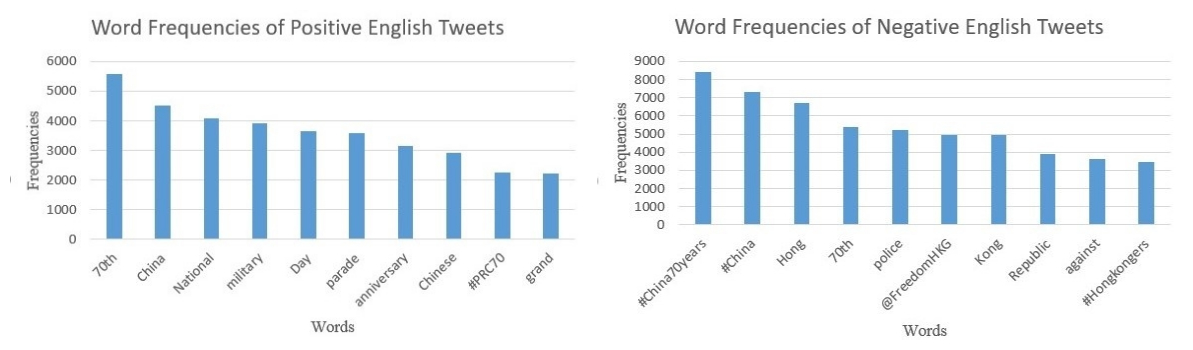}
\caption{The top 10 frequently used words of positive and negative English tweets}
\label{fig:13}       
\end{figure}

\begin{figure}
  \includegraphics[width=14cm,height=6cm]{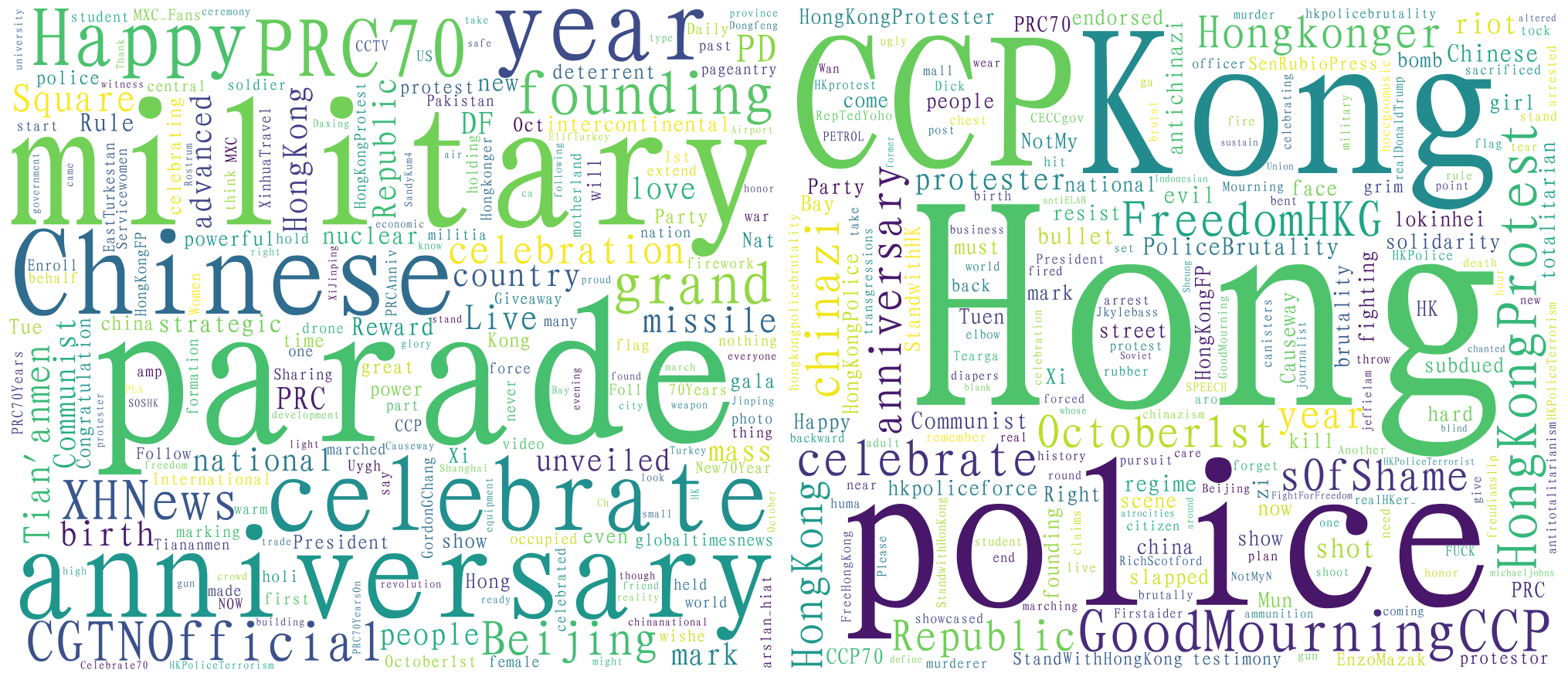}
\caption{The word clouds of positive and negative English tweets}
\label{fig:14}       
\end{figure}

\begin{figure}
  \includegraphics[width=16cm,height=5cm]{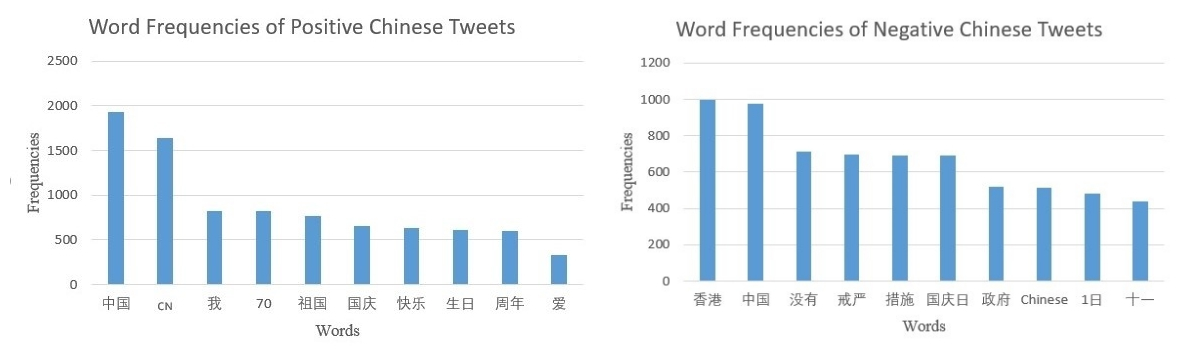}
\caption{The top 10 frequently used words of positive and negative Chinese tweets}
\label{fig:15}       
\end{figure}

\begin{figure}
  \includegraphics[width=14cm,height=6cm]{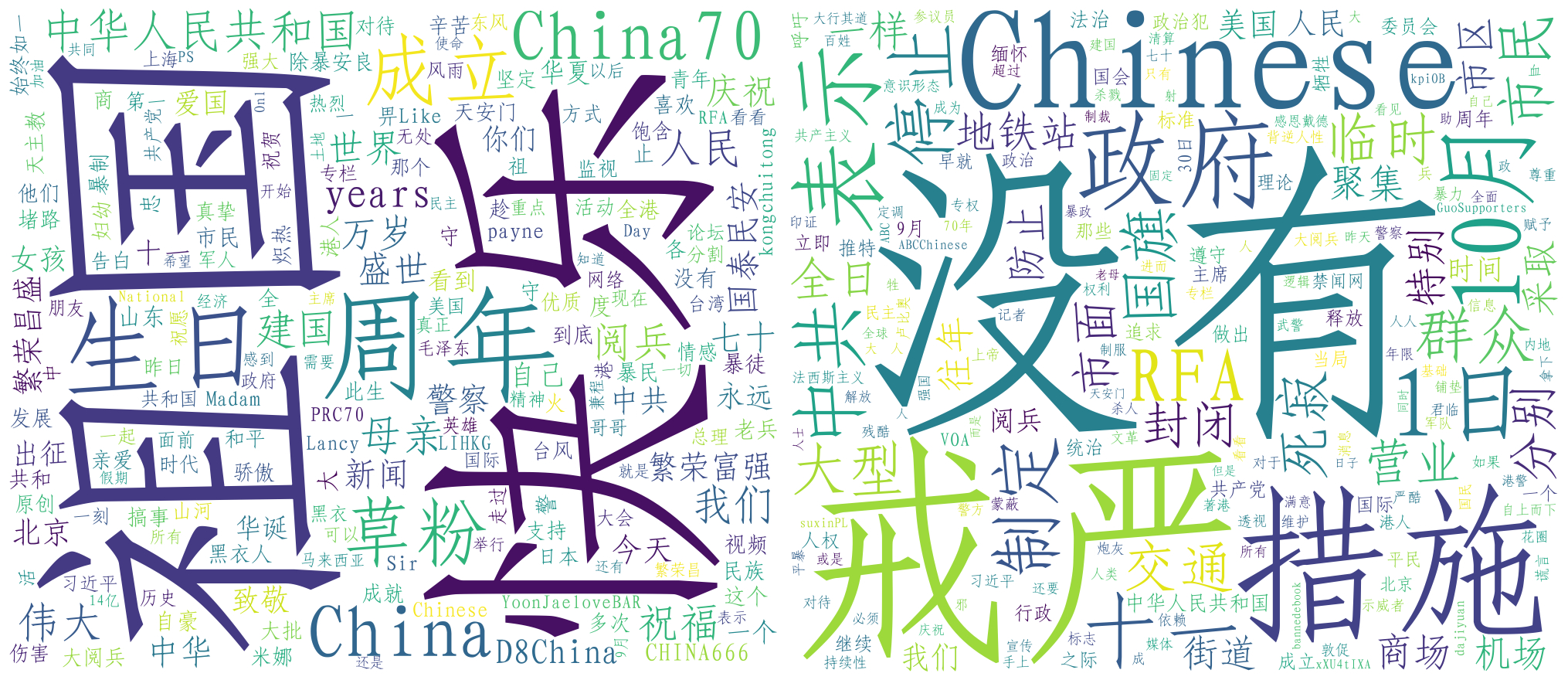}
\caption{The word clouds of positive and negative Chinese tweets}
\label{fig:16}       
\end{figure}

\section{Discussion, Conclusions and Future Works}
\label{sec:11}
In this descriptive study, specific Twitter data sent around the 2019 Chinese National Day was collected and their sentiments were analyzed with a hybrid technique. The temporal, spatial and lingual characteristics of the tweets were explored.

It can be found that the number of English and Chinese tweets sent every hour is generally in accordance with the celebration activities and the time zone of Twitter users. The highest peak of English tweets was the morning of October 1 in Europe, when European users were most likely to know the news. The second peak was around 11 o’clock on October 1 during the military parade. The most significant peak for Chinese tweets was around 10 o’clock on October 1 when the parade began and another prominent peak was around 20 o’clock when the firework show and celebration took place. As for the absolute sentiments, Figures 3 and 4 show that except for several hours from 10 o’clock to 12 o’clock on October 1 when a slightly positive attitude was expressed, the general sentiments of English tweets towards China were negative. Chinese tweets, on the contrary, were more positive to China, and the scores were above zero most of the time. The results of normalized sentiments confirmed this observation. Chinese tweets often have a positive view of China while English tweets became constantly negative since noon Oct 1.

For spatial characteristics, Figure 7 to 10 depict that people from China (including Hong Kong, Macau, and Taiwan, the same below), the United States, Pakistan, India, and Australia sent the most positive English tweets, while China, the United States, Uganda, and Japan produced the most negative English tweets towards China. As for tweets written in Chinese, China, the United States, Japan, and Thailand were the important positive sentiments providers, and the United States, China, Japan, Australia, and Canada were the most negative. Comparing the number of positive and negative tweets, the overall expressed sentiment of English tweets in most countries were positive, as is shown in Figure 11. The highest score, 89, was achieved by Pakistan. However, English tweets sent from Turkey, Uganda, China, Honduras, Poland, Venezuela, and Germany were in general negative. Meanwhile, the sentiments expressed by Chinese tweets depict a different image. Cambodia, Canada, the United States, Germany, and Australia became the only nations that hold generally negative views towards China. China, Russia, France, the United Kingdom, and Pakistan were the most positive nations. Interestingly, for countries like the United States, Canada, and Australia, English tweets were more positive than Chinese tweets.

The linguistic features reveal the different focus points of Twitter users whose sentiments to China were conflicting. The positive side focused more on the celebration activities and the military parade, and expressed all kinds of good wishes for China. On the other hand, the negative discourses were mainly connected with the Hong Kong issue.

Although little theoretical improvement or causal inference was achieved in this study, the phenomenon disclosed through the analysis is still worthy of notice and further research. One of the intuitive but significant findings is that English tweets as a whole were more negative towards China, and countries that enjoy relatively better relationship with China tend to hold a more positive view, which indicates a certain consistency between the online opinion and the official relations. This indicates notable discrepancies among the perception of China in different countries, which not only reflect the official attitudes and the public opinions, but also have a realistic impact on China. For example, cyber security and cyber-attacks, a hot topic in Sino-US relations in recent years, are gradually entering the field of diplomacy and international relations.[23] However, according to Kumar and Carley, the probability of experiencing cyber-attacks increases by up to 27\% if the country is being viewed negatively by other nations[24, 25], and this situation can in turn increase tensions. In addition, online public opinion may influence a country’s future foreign policy towards China. Issues and opinions emphasized by social media would be regarded by the public as important. Thus, social media focus could be transferred into public and government agenda.[26, 27] Georgiadou et al. have shown the importance and potential of utilizing social media sentiment analysis to enhancing decision-making during international negotiations such as the UK-EU Brexit negotiation.[28] Negative sentiments can also play a role in the exacerbating and accelerating of the politicization and securitization of China-related issues by providing suitable social background, which may finally influence the bilateral relationships.[29, 30] Therefore, online attitudes towards China are indicative and non-negligible.

Further studies are supposed to both refine the techniques used to yield more fine-grained sentiments results as well as to improve the experiment design. For example, the data collecting time needs to be prolonged to include more events, especially events with small probability and cannot be predicted.



\begin{acknowledgements}
We are grateful for Xin Yin, Hu Lingrui, Liu Yu, and an anonymous student at TBSI for their work in labeling our 2,000 tweets training set. We would like to give our special thanks to Dr. Wan Zuofang at the National Institute of Education Sciences and Professor Hu Yue at the Department of Political Science, Tsinghua University. Our discussion with them has inspired several parts of this research.
\end{acknowledgements}

%
%


\begin{thebibliography}{}
%
%
\bibitem{RefJ}
X. Han, E. Choi, and C. Tan, "No Permanent Friends or Enemies: Tracking Relationships between Nations from News," \textit{arXiv preprint arXiv:1904.08950}, 2019.
\bibitem{RefJ}
R. Reardon and N. Choucri, "The role of cyberspace in international relations: A view of the literature," in \textit{ISA Annual Convention, San Diego, CA}, 2012, vol. 1.
\bibitem{RefJ}
S. Sun, C. Luo, and J. Chen, "A review of natural language processing techniques for opinion mining systems," \textit{Information Fusion}, vol. 36, pp. 10-25, 2017.
\bibitem{RefJ}
N. Tsirakis, V. Poulopoulos, P. Tsantilas, and I. Varlamis, "Large scale opinion mining for social, news and blog data," \textit{Journal of Systems and Software}, vol. 127, pp. 237-248, 2017.
\bibitem{RefJ}
A. Tsakalidis, N. Aletras, A. I. Cristea, and M. Liakata, "Nowcasting the stance of social media users in a sudden vote: The case of the Greek Referendum," in \textit{Proceedings of the 27th ACM International Conference on Information and Knowledge Management}, 2018, pp. 367-376. 
\bibitem{RefJ}
A. Jungherr, H. Schoen, O. Posegga, and P. Jürgens, "Digital trace data in the study of public opinion: An indicator of attention toward politics rather than political support," \textit{Social Science Computer Review}, vol. 35, no. 3, pp. 336-356, 2017.
\bibitem{RefJ}
A. Bovet, F. Morone, and H. A. Makse, "Validation of Twitter opinion trends with national polling aggregates: Hillary Clinton vs Donald Trump," \textit{Scientific reports}, vol. 8, no. 1, pp. 1-16, 2018.
\bibitem{RefJ}
J. C. A. D. Lopez, S. Collignon-Delmar, K. Benoit, and A. Matsuo, "Predicting the brexit vote by tracking and classifying public opinion using twitter data," \textit{Statistics, Politics and Policy}, vol. 8, no. 1, pp. 85-104, 2017.
\bibitem{RefJ}
R. Gull, U. Shoaib, S. Rasheed, W. Abid, and B. Zahoor, "Pre processing of twitter's data for opinion mining in political context," \textit{Procedia Computer Science}, vol. 96, pp. 1560-1570, 2016.
\bibitem{RefJ}
W. Gong, Y. Zhang, and H. Cai, "Haiwai zimeiti zhong shehua yuqing chuanbo jizhi de dashuju fenxi——jiyu Reddit pingtai de hailiang yuqing xinxi [Big Data Analysis of China-related Public Opinion Dissemination Mechanism in Overseas Social Media: Massive Public Opinion Information Based on Reddit Platform](In Chinese)," \textit{Academic Forum}, vol. 2017, no. 3, pp. 21-31, 2017.
\bibitem{RefJ}
A. A. Jamal, R. O. Keohane, D. Romney, and D. Tingley, "Anti-Americanism and anti-interventionism in Arabic Twitter discourses," \textit{Perspectives on Politics}, vol. 13, no. 1, pp. 55-73, 2015.
\bibitem{RefJ}
Y. Guan, D. Tingley, D. Romney, A. Jamal, and R. Keohane, "Chinese views of the United States: evidence from Weibo," \textit{International Relations of the Asia-Pacific}, vol. 00, pp. 1-30, 2018.
\bibitem{RefJ}
M. Di Giovanni, M. Brambilla, S. Ceri, F. Daniel, and G. Ramponi, "Content-based Classification of Political Inclinations of Twitter Users," in \textit{2018 IEEE International Conference on Big Data (Big Data)}, 2018: IEEE, pp. 4321-4327. 
\bibitem{RefJ}
G. A. Barnett et al., "Measuring international relations in social media conversations," \textit{Government Information Quarterly}, vol. 34, no. 1, pp. 37–44, 2017.
\bibitem{RefJ}
S. H. Kumar, J. Pujara, L. Getoor, D. Mares, D. Gupta, and E. Riloff, "Unsupervised models for predicting strategic relations between organizations," in \textit{Proceedings of the 2016 IEEE/ACM International Conference on Advances in Social Networks Analysis and Mining}, 2016: IEEE Press, pp. 711-718. 
\bibitem{RefJ}
N. Chambers et al., "Identifying political sentiment between nation states with social media," in \textit{Proceedings of the 2015 Conference on Empirical Methods in Natural Language Processing}, 2015, pp. 65-75. 
\bibitem{RefJ}
J. Brandt et al., "Identifying social media user demographics and topic diversity with computational social science: a case study of a major international policy forum," \textit{Journal of Computational Social Science}, pp. 1-22.
\bibitem{RefJ}
H. Rashkin, E. Bell, Y. Choi, and S. Volkova, "Multilingual connotation frames: A case study on social media for targeted sentiment analysis and forecast," in \textit{Proceedings of the 55th Annual Meeting of the Association for Computational Linguistics (Volume 2: Short Papers)}, 2017, pp. 459-464.
\bibitem{RefJ}Q. Fang, C. Xu, J. Sang, M. S. Hossain, and G. Muhammad, "Word-of-mouth understanding: Entity-centric multimodal aspect-opinion mining in social media," \textit{IEEE Transactions on Multimedia}, vol. 17, no. 12, pp. 2281-2296, 2015.
\bibitem{RefJ}
V. Peña-Araya, M. Quezada, B. Poblete, and D. Parra, "Gaining historical and international relations insights from social media: spatio-temporal real-world news analysis using Twitter," \textit{Epj Data Science}, vol. 6, no. 1, p. 25, 2017.
\bibitem{RefJ}
S. Fernando, J. A. D. López, O. Şerban, J. Gómez-Romero, M. Molina-Solana, and Y. Guo, "Towards a large-scale twitter observatory for political events," \textit{Future Generation Computer Systems}, 2019.
\bibitem{RefJ}
X. Tannier, "NLP-driven data journalism: Time-aware mining and visualization of international alliances," in \textit{Proceedings of the 2016 IJCAI Workshop on Natural Language Processing meets Journalism}, 2016, pp. 52-56. 
\bibitem{RefJ}
A. Barrinha and T. Renard, "Cyber-diplomacy: the making of an international society in the digital age," \textit{Global Affairs}, vol. 3, no. 4-5, pp. 353-364, 2017.
\bibitem{RefJ}
S. Kumar and K. M. Carley, "Understanding DDoS cyber-attacks using social media analytics," in \textit{2016 IEEE Conference on Intelligence and Security Informatics (ISI)}, 2016: IEEE, pp. 231-236. 
\bibitem{RefJ}
S. Kumar and K. M. Carley, "Approaches to understanding the motivations behind cyber attacks," in \textit{2016 IEEE Conference on Intelligence and Security Informatics (ISI)}, 2016: IEEE, pp. 307-309. 
\bibitem{RefJ}
J. T. Feezell, "Agenda setting through social media: The importance of incidental news exposure and social filtering in the digital era," \textit{Political Research Quarterly}, vol. 71, no. 2, pp. 482-494, 2018.
\bibitem{RefJ}
G. King, B. Schneer, and A. White, "How the news media activate public expression and influence national agendas," \textit{Science}, vol. 358, no. 6364, pp. 776-780, 2017.
\bibitem{RefJ}
E. Georgiadou, S. Angelopoulos, and H. Drake, "Big data analytics and international negotiations: Sentiment analysis of Brexit negotiating outcomes," \textit{International Journal of Information Management}, p. 102048, 2019.
\bibitem{RefJ}
T. Balzacq, "The three faces of securitization: Political agency, audience and context," \textit{European journal of international relations}, vol. 11, no. 2, pp. 171-201, 2005.
\bibitem{RefJ}
S. Koschut, T. H. Hall, R. Wolf, T. Solomon, E. Hutchison, and R. Bleiker, "Discourse and emotions in international relations," \textit{International Studies Review}, vol. 19, no. 3, pp. 481-508, 2017.
\end{thebibliography}


\end{document}